\documentclass[prl,preprint,showpacs,preprintnumbers,superscriptaddress,nofootinbib]{revtex4}
\usepackage{amsmath}

\usepackage{graphicx}

\usepackage{graphicx}% Include figure files
\usepackage{dcolumn}% Align table columns on decimal point
\usepackage{bm}% bold math
\usepackage{amsmath}
\usepackage{amssymb}
\usepackage{url}
\usepackage[normalem]{ulem}

\newcommand{\beq}{\begin{equation}}
\newcommand{\eeq}{\end{equation}}
\newcommand{\bea}{\begin{eqnarray}}
\newcommand{\eea}{\end{eqnarray}}
\newcommand{\real}{{\sf I}\kern-.12em{\sf R}}
\newcommand{\comp}{{\sf I}\kern-.50em{\sf C}}
\newcommand{\unity}{{\sf I}\kern-.54em{\sf 1}}

\newcommand{\PS}{d({\rm P.S.})}

\def\spose#1{\hbox to 0pt{#1\hss}}
\def\ltapprox{\mathrel{\spose{\lower 3pt\hbox{$\mathchar"218$}}
 \raise 2.0pt\hbox{$\mathchar"13C$}}}

\begin{document}

\title{High-Energy Resummation in Di-hadron Production at the LHC}
\author{Francesco G. Celiberto}
%\email{francescogiovanni.celiberto@fis.unical.it}
\affiliation{Dipartimento di Fisica dell'Universit\`a della Calabria \\
I-87036 Arcavacata di Rende, Cosenza, Italy}
\affiliation{INFN - Gruppo collegato di Cosenza, I-87036 Arcavacata di Rende,
Cosenza, Italy}
\author{Dmitry Yu. Ivanov}
\affiliation{Sobolev Institute of Mathematics, 630090 Novosibirsk, Russia}
\affiliation{Novosibirsk State University, 630090 Novosibirsk, Russia}
%\email{d-ivanov@math.nsc.ru}
\author{Beatrice Murdaca}
%\email{beatrice.murdaca@fis.unical.it}
\affiliation{INFN - Gruppo collegato di Cosenza, I-87036 Arcavacata di Rende,
Cosenza, Italy}
\author{Alessandro Papa}
%\email{alessandro.papa@fis.unical.it}
\affiliation{Dipartimento di Fisica dell'Universit\`a della Calabria \\
I-87036 Arcavacata di Rende, Cosenza, Italy}
\affiliation{INFN - Gruppo collegato di Cosenza, I-87036 Arcavacata di Rende,
Cosenza, Italy}

\begin{abstract}
We propose to study at the Large Hadron Collider (LHC) the inclusive 
production  of a pair of hadrons  (a ``di-hadron'' system)  
in a kinematics where two detected hadrons with high transverse 
momenta are separated by a large interval of rapidity. This process has much in 
common with the widely discussed Mueller-Navelet jet production and can also 
be used to access the dynamics of hard proton-parton interactions in the Regge 
limit. For both processes  large contributions enhanced by logarithms of 
energy can be resummed in perturbation theory  within the 
Balitsky-Fadin-Kuraev-Lipatov (BFKL) formalism with next-to-leading 
logarithmic accuracy (NLA). The experimental study of di-hadron production 
would provide with an additional clear channel to test the BFKL dynamics. 
We present here the first theoretical predictions for cross sections and 
azimuthal angle correlations of the di-hadrons produced  with  LHC 
kinematics.

\end{abstract}
\maketitle

\section{Introduction}

The record energy of proton-proton collisions and the high luminosity of LHC 
provide us with a unique opportunity to study the dynamics of strong 
interactions in a kinematic range  so far unexplored. The production of 
two, the most forward and backward, jets, separated by a large interval of 
rapidity, was proposed by Mueller and Navelet~\cite{Mueller:1986ey} as a tool 
to access the dynamics of semihard parton interactions at a hadron collider. 
In theory such processes are described using the BFKL 
method~\cite{Fadin:1975cbKuraev:1976geKuraev:1977fsBalitsky:1978ic}, which 
allows to resum  to all orders  the leading (LLA) and the next-to-leading 
terms (NLA) of the QCD perturbative series that are enhanced by powers of 
large energy logarithms.
For the Mueller-Navelet jet production, the BFKL resummation with NLA accuracy  
relies  on the combination of two ingredients: the NLA Green's function 
of the BFKL equation~\cite{Fadin:1998py,Ciafaloni:1998gs} and the NLA jet 
vertices~\cite{Bartels:2001ge,Bartels:2002yj,Caporale:2011cc,Ivanov:2012ms,Colferai:2015zfa}.
In~\cite{Colferai:2010wu,Caporale:2012ih,Ducloue:2013hia,Ducloue:2013bva,Caporale:2013uva,Ducloue:2014koa,Caporale:2014gpa,Ducloue:2015jba,Caporale:2015uva,Celiberto:2015yba,Celiberto:2016ygs}, NLA BFKL calculations of the cross sections 
for the Mueller-Navelet jet process and also predictions 
for the jet azimuthal angle correlations, observables earlier suggested
in~\cite{DelDuca:1993mn,Stirling:1994he}, can be found. 
Recently~\cite{Khachatryan:2016udy}, the first measurements of the azimuthal 
correlation of the Mueller-Navelet jets at LHC were presented by the CMS 
Collaboration at $\sqrt{s}=7$~{TeV}. Further experimental studies of 
the Mueller-Navelet jets at higher LHC energies and larger rapidity 
separations are expected.  

The important task of revealing the dynamical mechanisms behind
partonic interactions in the Regge limit, $s\gg |t|$, by the comparison
of theory predictions with data, can be better accomplished if some other
observables, sensitive to the BFKL dynamics, are considered in the context of 
the LHC physics program. An interesting option, the detection of three 
jets and four jets, well separated in rapidity from each other, was recently 
suggested in~\cite{Caporale:2015vya} and in~\cite{Caporale:2015int}.

In this letter we want to suggest a novel possibility, {\it i.e.} 
the inclusive di-hadron production 
\begin{eqnarray}
\label{process}
{\rm p}(p_1) + {\rm p}(p_2) \to {\rm h}_1(k_1) + {\rm h}_2(k_2)+ {\rm X} \;,
\end{eqnarray}
when the two detected hadrons have high transverse momenta and are separated 
by a large interval of rapidity. For this process, similarly to the  
Mueller-Navelet jet production, the BFKL resummation in the NLA is feasible,
since the necessary  item  beyond the NLA BFKL Green's functions, {\it i.e.}
the vertex describing the production of an identified hadron, was obtained
with NLA in~\cite{hadrons}. 
It  was  shown there that, after renormalization of the QCD coupling 
and  the  ensuing removal of the ultraviolet divergences, soft and virtual 
infrared divergences cancel each other, whereas the  surviving  infrared 
collinear ones are compensated by the collinear counterterms related with 
the renormalization of parton densities (PDFs) for the initial proton and 
parton fragmentation functions (FFs) describing the detected hadron in the 
final state within collinear factorization. 
All the theoretical requisites are  thus  fulfilled to write down 
infrared-safe NLA predictions, thus making of this process an additional 
clear channel to test the BFKL dynamics at the LHC. The fact that hadrons 
can be detected at the LHC at much smaller values of the transverse
momentum than jets, allows to explore a  kinematic range outside the
reach of the Mueller-Navelet channel, so that the reaction~(\ref{process}) 
can be considered complementary to Mueller-Navelet jet production, although
sharing with it the theoretical framework.

We will give below the very first predictions for the cross sections and 
azimuthal angle correlations of the process~(\ref{process}).  We will
limit ourselves to present the main formulas, so to guarantee the 
reproducibility of our results, and postpone a more detailed account about
their derivation to a later publication. 

It is known that the inclusion of NLA terms makes a very large effect on the 
theory predictions for the Mueller-Navelet jet cross sections and the jet 
azimuthal angle distributions. Similar features are expected also for our case 
of inclusive di-hadron production. This results in a large dependence of 
predictions on the choice of  the renormalization scale $\mu_R$ 
and the factorization scale $\mu_F$. Here we will take  them equal, 
$\mu_R=\mu_F$, and adopt the Brodsky-Lepage-Mackenzie (BLM) 
scheme~\cite{Brodsky:1982gc} for the renormalization scale setting. 
In BLM the renormalization scale ambiguity is eliminated by 
absorbing the non-conformal, proportional to the QCD $\beta_0$-function,
terms into the running coupling. Such approach was successfully used, first 
in~\cite{Ducloue:2013bva}, for a satisfactory description of the LHC data on 
the azimuthal correlations of Mueller-Navelet jets~\cite{Khachatryan:2016udy}, 
obtained  by the CMS collaboration. 

\section{BFKL with BLM optimization}

We consider the production, in high-energy proton-proton collisions, of a pair 
of identified hadrons with large transverse momenta, $\vec k_1^2\sim 
\vec k_2^2 \gg \Lambda^2_{QCD}$ and large separation in rapidity.

In collinear factorization we neglect power-suppressed contributions, therefore 
the proton mass can be taken vanishing and the Sudakov vectors can be chosen
to coincide with the proton momenta $p_1$ and $p_2$, satisfying $p^2_1= p^2_2=0$
and  $2 \, p_1\cdot p_2 = s$. 
Then, the momentum of each identified hadron can be decomposed as 
\beq
k_{1,2}= \alpha_{1,2} p_{1,2} + \frac{\vec k_{1,2}^2}{\alpha_{1,2} s}p_{2,1}
+k_{1,2 \perp} \ , \quad k_{1,2 \perp}^2=-\vec k_{1,2}^2  \ .
\label{sudakov}
\eeq
In the center-of-mass system, the longitudinal fractions $\alpha_{1,2}$ are
related to the hadron rapidities by $y_1=\frac{1}{2}\ln\frac{\alpha_1^2 s}
{\vec k_1^2}$ and $y_2=\frac{1}{2}\ln\frac{\vec k_2^2}{\alpha_2^2 s}$,
which imply $dy_1=\frac{d\alpha_1}{\alpha_1}$ and $dy_2=-\frac{d\alpha_2}
{\alpha_2}$, if the space part of the four-vector $p_{1}$ is taken positive.
The differential cross section of the process~(\ref{process}) can be written 
as follows: 
\beq
\frac{d\sigma}{dy_1dy_2\, d|\vec k_1| \, d|\vec k_2|d\phi_1 d\phi_2}
=
\frac{1}{(2\pi)^2}
\left[
{\cal C}_0+\sum_{n=1}^\infty  2\cos (n\phi ){\cal C}_n \right] \;,
\eeq
where $\phi=\phi_1-\phi_2-\pi$, with $\phi_{1,2}$  the two hadrons'
azimuthal angles, while $y_{1,2}$ and $\vec k_{1,2}$ are their
rapidities and transverse momenta, respectively. The eigenvalues of the
kernel of the BFKL equation and the expressions for the hadron vertices 
are needed to calculate this cross section. 
In LLA the BFKL eigenvalues, parameterized by the continuous $\nu$ variable 
and the integer conformal spin parameter $n$, read   
$$
\chi\left(n,\nu\right)=2\psi\left(1\right)-\psi\left(\frac{n}{2}
+\frac{1}{2}+i\nu \right)-\psi\left(\frac{n}{2}+\frac{1}{2}-i\nu \right),
$$
and the LLA hadron vertices,
\beq
\label{c1}
c_1(n,\nu,|\vec k_1|,\alpha_1) = \frac{4}{3}
(\vec k_1^2)^{i\nu-1/2}\,\int_{\alpha_1}^1\frac{dx}{x}
\left( \frac{x}{\alpha_1}\right)^{2 i\nu-1} 
\left[\frac{C_A}{C_F}f_g(x)D_g^h\left(\frac{\alpha_1}{x}\right)
+\sum_{a=q,\bar q}f_a(x)D_a^h\left(\frac{\alpha_1}{x}\right)\right] ,
\eeq
\beq
\label{c2}
c_2(n,\nu,|\vec k_2|,\alpha_2)=\biggl[c_1(n,\nu,|\vec k_2|,\alpha_2)\biggr]^* \;,
\eeq
are given as an integral in the parton fraction $x$, containing  the
PDFs of the gluon and of the different quark/antiquark flavors in the proton, 
and  the  FFs of the detected hadron  (for more details, 
see~\cite{hadrons}). 
It is known~\cite{Brodsky:1996sgBrodsky:1997sdBrodsky:1998knBrodsky:2002ka}, 
that in the BLM approach applied to semihard processes, we need to perform a 
finite renormalization from the $\overline{\rm MS}$ to the physical MOM scheme:
\beq{}
\alpha_s^{\overline{\rm MS}}=\alpha_s^{\rm MOM}\left(1+\frac{\alpha_s^{\rm MOM}}{\pi}T
\right)\;,
\eeq
with $T=T^{\beta}+T^{\rm conf}$,
\beq{}
T^{\beta}=-\frac{\beta_0}{2}\left( 1+\frac{2}{3}I \right)\, ,
\eeq
\[ 
T^{\rm conf}= \frac{3}{8}\left[ \frac{17}{2}I +\frac{3}{2}\left(I-1\right)\xi
+\left( 1-\frac{1}{3}I\right)\xi^2-\frac{1}{6}\xi^3 \right] \;,
\]
where $I=-2\int_0^1dx\frac{\ln\left(x\right)}{x^2-x+1}\simeq2.3439$ and $\xi$
is the gauge parameter of the MOM scheme, fixed at zero in the following.
The optimal scale $\mu_R^{\rm BLM}$ is the value of $\mu_R$ that makes 
 the $\beta_0$-dependent part in the expression for the observable
of interest  vanish.  
In~\cite{Caporale:2015uva} some of us showed that terms proportional 
to the QCD $\beta_0$-function are present not only in the NLA BFKL kernel, 
but also in the expressions for the NLA vertices (called ``impact factors''
in the BFKL jargon). This leads 
to a non-universality of the BLM scale and to its dependence on the 
energy of the process. It was also found~\cite{Caporale:2015uva} that 
contributions proportional to the NLA impact factors are universally expressed 
in terms of the LLA impact factors of the considered process, through
the function $f\left(\nu\right)$, defined as follows: 
\beq{}
\label{nu}
i\frac{d}{d\nu}\ln\left(\frac{c_1}{c_2}\right)\equiv 2 \left[f(\nu)
-\ln\left(|\vec k_1||\vec k_2|\right)\right]\ .
\eeq
Finally, the condition for the BLM scale setting  was found to be 
%\begin{widetext}
\[
C^{\beta}_n
\propto \!\!\int\PS \!\! 
\int\limits^{\infty}_{-\infty} \!\!d\nu\,e^{Y \bar \alpha^{\rm MOM}_s(\mu^{\rm BLM}_R)\chi(n,\nu)}
%\left(\alpha^{\rm MOM}_s (\mu^{\rm BLM}_R)\right)^3
c_1(n,\nu)c_2(n,\nu)
%\frac{\beta_0}{2 N_c}
\left[\frac{5}{3}
+\ln \frac{(\mu^{\rm BLM}_R)^2}{|\vec k_1|
|\vec k_2|} +f(\nu)-2\left( 1+\frac{2}{3}I \right)
\right.
\]
\beq{}
\label{beta0}
\left.
+\bar \alpha^{\rm MOM}_s(\mu^{\rm BLM}_R) Y \: \frac{\chi(n,\nu)}{2}
\left(-\frac{\chi(n,\nu)}{2}+\frac{5}{3}+\ln \frac{(\mu^{\rm BLM}_R)^2}{|\vec k_1|
|\vec k_2|}
+f(\nu)-2\left( 1+\frac{2}{3}I \right)\right)\right]=0 \, ,
\eeq
%\end{widetext}
where $\bar \alpha_s={3\alpha_s/\pi }$; the first term in the r.h.s. 
of~(\ref{beta0}) originates from the NLA corrections to the hadron vertices 
and the second one from the NLA part of the kernel. Here $Y$ is the rapidity
separation of two detected hadrons, $Y=y_1-y_2$. 
We consider the \emph{coefficients} integrated over the phase space for 
two final state hadrons,  
\beq
\label{Cm_int}
C_n= \int\PS  \, {\cal C}_n \left(y_1,y_2,k_1,k_2 \right)\, ,
\eeq
where
\beq
\int\PS= \int_{k_{1,\rm min}}^{\infty}d|\vec k_1|
\int_{k_{2,\rm min}}^{\infty}d|\vec k_2|
\int_{y_{1,\rm min}}^{y_{1,\rm max}}dy_1
\int_{y_{2,\rm min}}^{y_{2,\rm max}}dy_2 \, \delta\left(y_1-y_2-Y\right) \, .
\eeq
For the  integrations over rapidities  we use the limits, 
$y_{1,\rm min}=-y_{2,\rm max}=-2.4$, $y_{1,\rm max}=-y_{2,\rm min}=2.4$, that are 
typical for the identified hadron detection at LHC.
As minimum transverse momenta we choose $k_{1,\rm min}=k_{2,\rm min}=5$~GeV,
 which are also realistic values for the LHC. We observe that 
the minimum transverse momentum in the CMS analysis~\cite{Khachatryan:2016udy} 
of Mueller-Navelet jet production is much larger, $k_{\rm jet,\rm min}=35$~GeV. 
In our calculations we use the PDF set MSTW 2008 NLO~\cite{Martin:2009iq} 
with two different NLO parameterizations for hadron FFs:  
AKK~\cite{Albino:2008fy} and HKNS~\cite{Hirai:2007cx}.  We considered
also the DSS parametrization~\cite{DSS}, but do not show the related
results here, since they would be hardly distinguishable from those with 
the HKNS parametrization. 
In the results presented below we sum over the production of charged light 
hadrons: $\pi^{\pm}, K^{\pm}, p,\bar p$.

%\begin{widetext}
\begin{figure*}[htb]
\centering
\includegraphics[scale=0.40,clip]{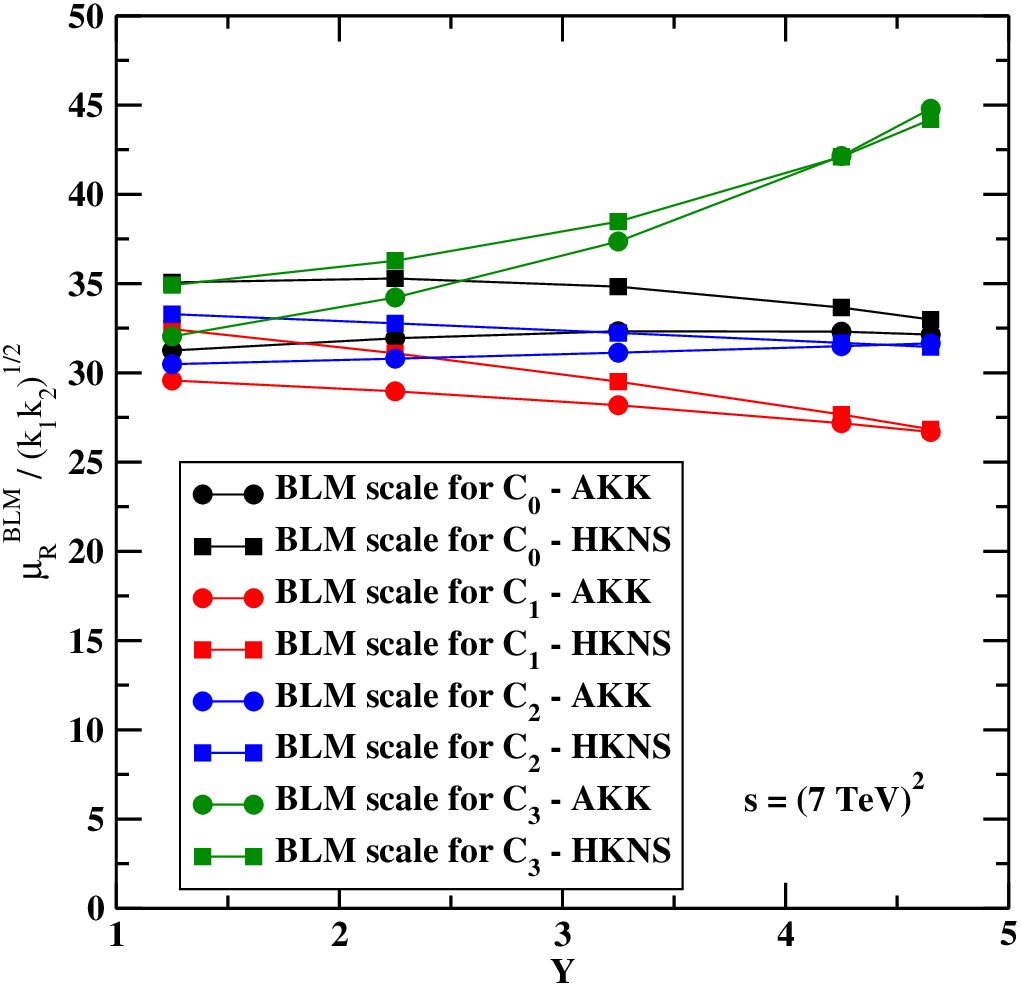} \ \ \
\includegraphics[scale=0.40,clip]{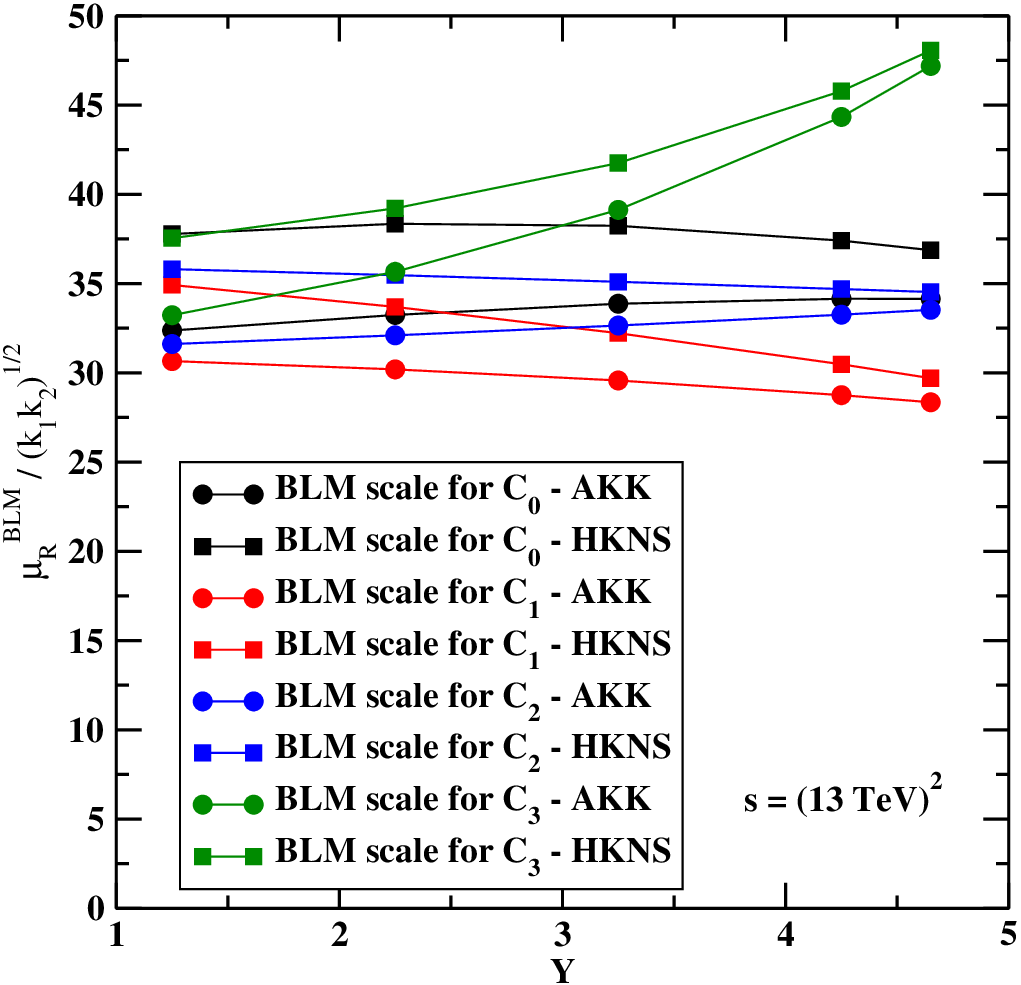}
\caption[]{BLM scales for the di-hadron production 
 {\it versus} the rapidity interval $Y$  for the two parametrizations
of FFs and for all the observables considered in this work. }
\label{scales}
\end{figure*}
%\end{widetext}

\begin{figure*}[htb]
\centering
\includegraphics[scale=0.40,clip]{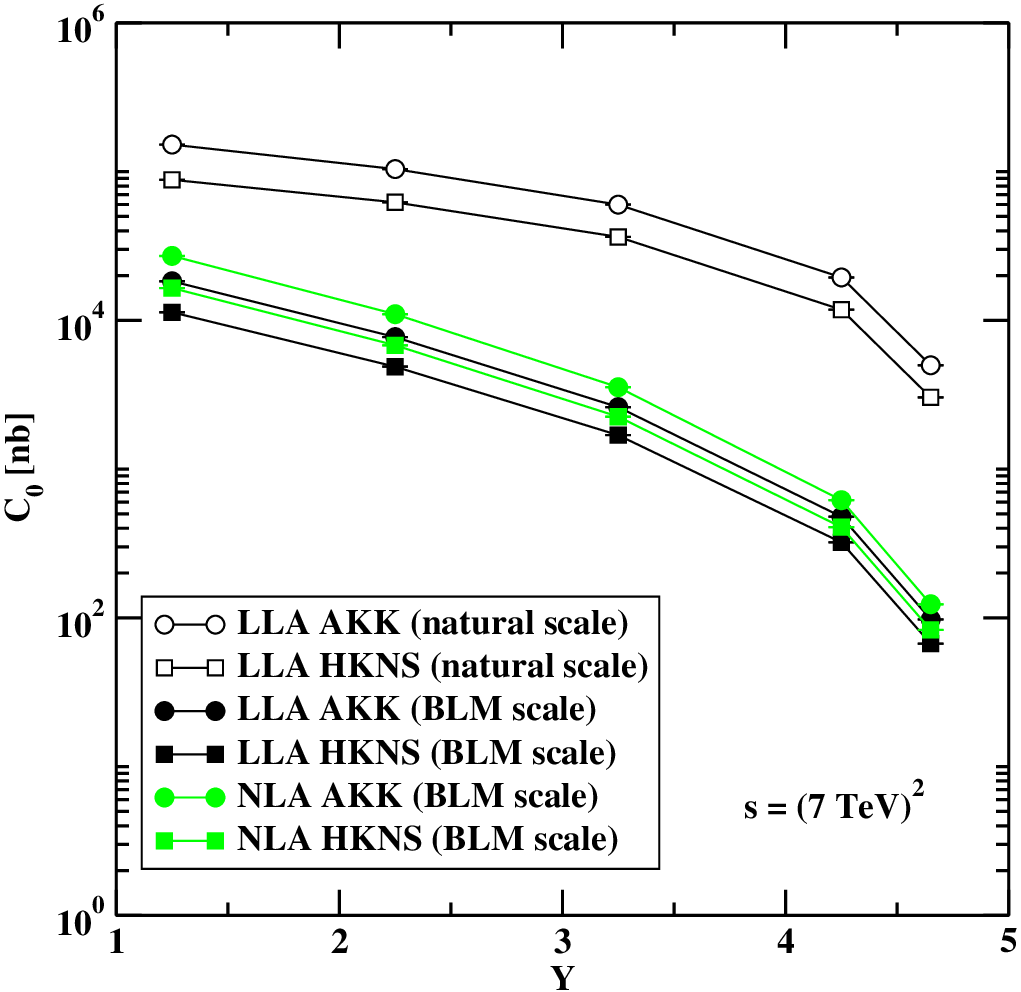} \ \ \
\includegraphics[scale=0.40,clip]{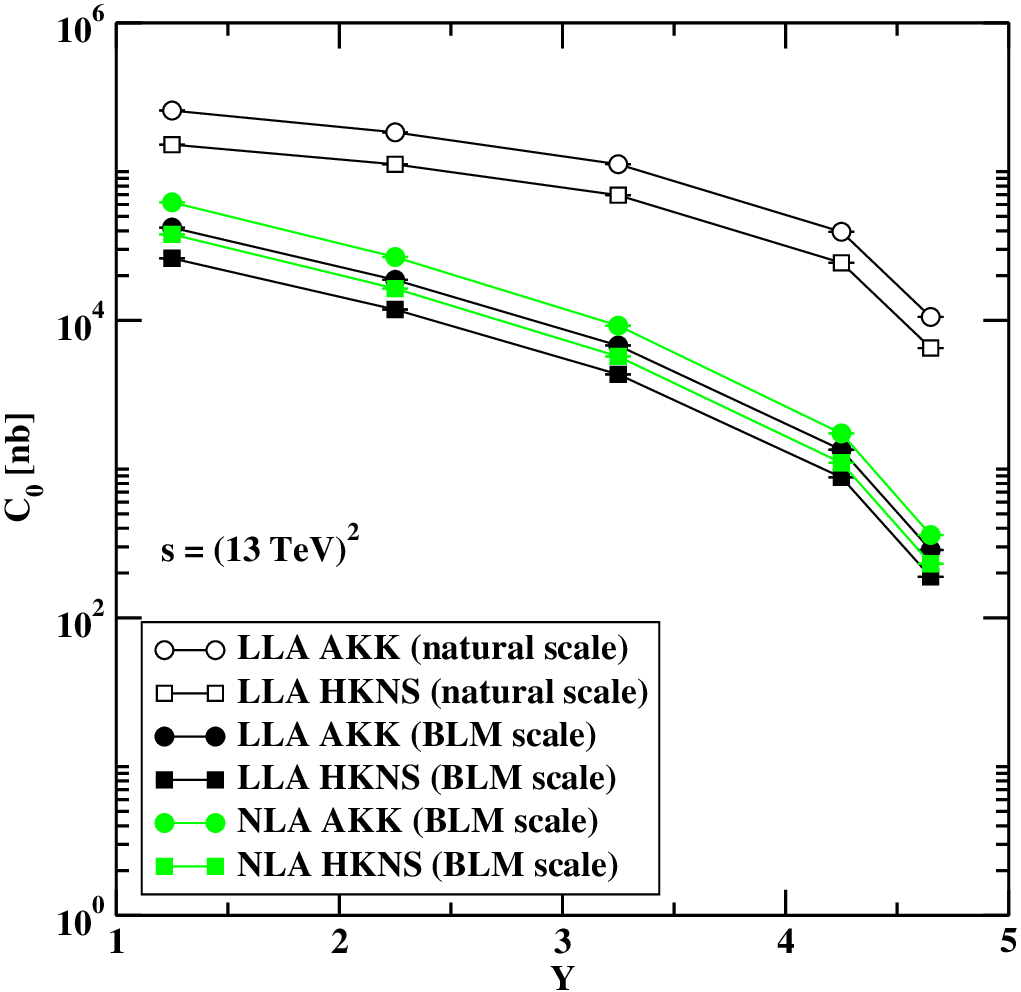}
\caption[]{Cross sections of the di-hadron production at LHC
{\it versus} the rapidity interval $Y$  for the two parametrizations
of FFs considered in this work: a) $\sqrt{s}$=7 TeV, 
b) $\sqrt{s}$=13 TeV. See the text for the definition of ``natural'' 
and ``BLM'' scales.}
\label{cross_sections}
\end{figure*}

\begin{figure*}[htb]
\centering
\includegraphics[scale=0.40,clip]{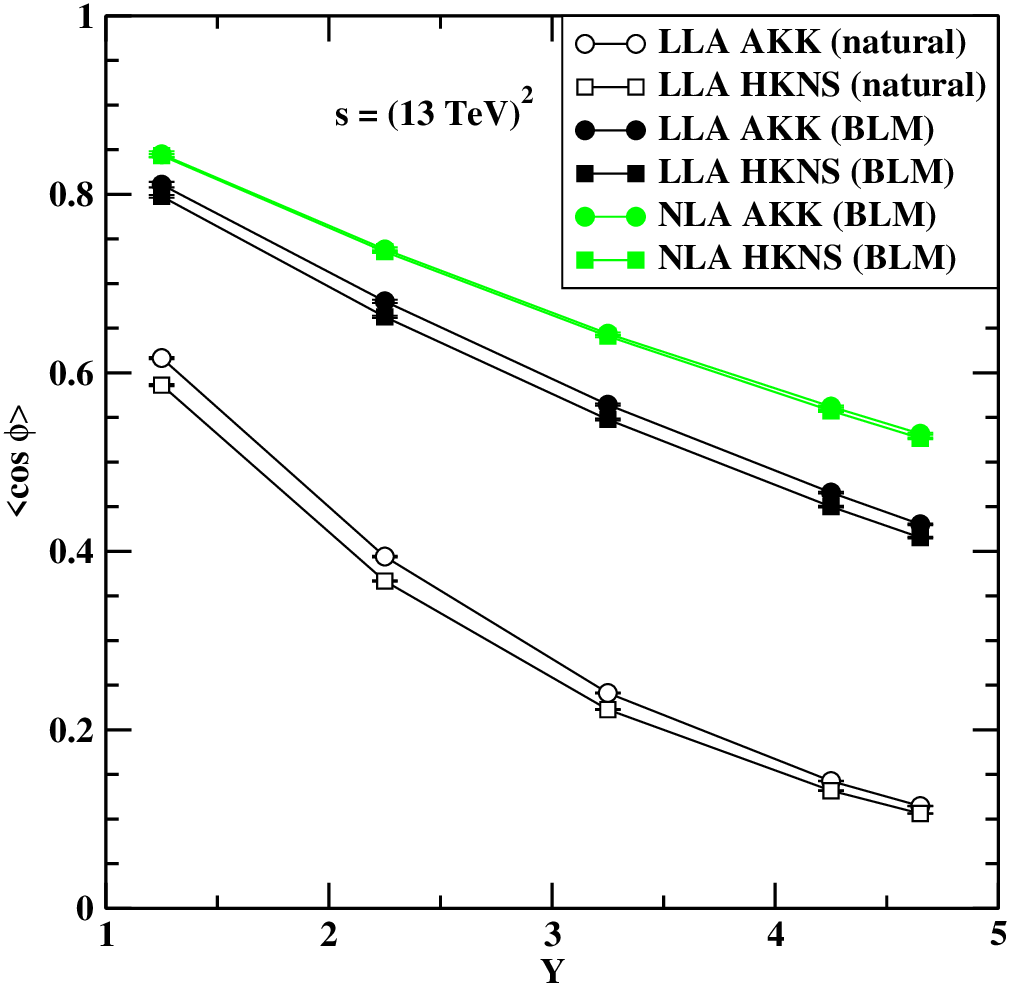} \ \ \
\includegraphics[scale=0.40,clip]{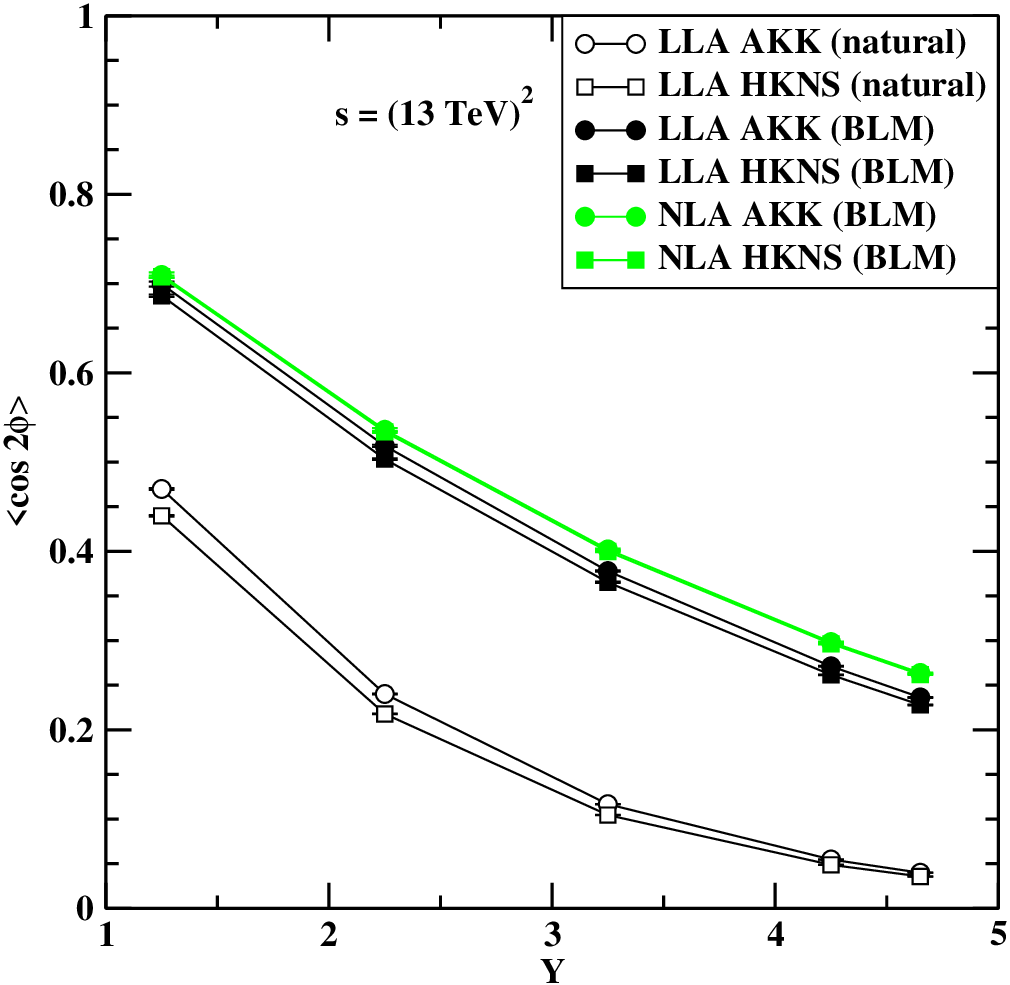}

\includegraphics[scale=0.40,clip]{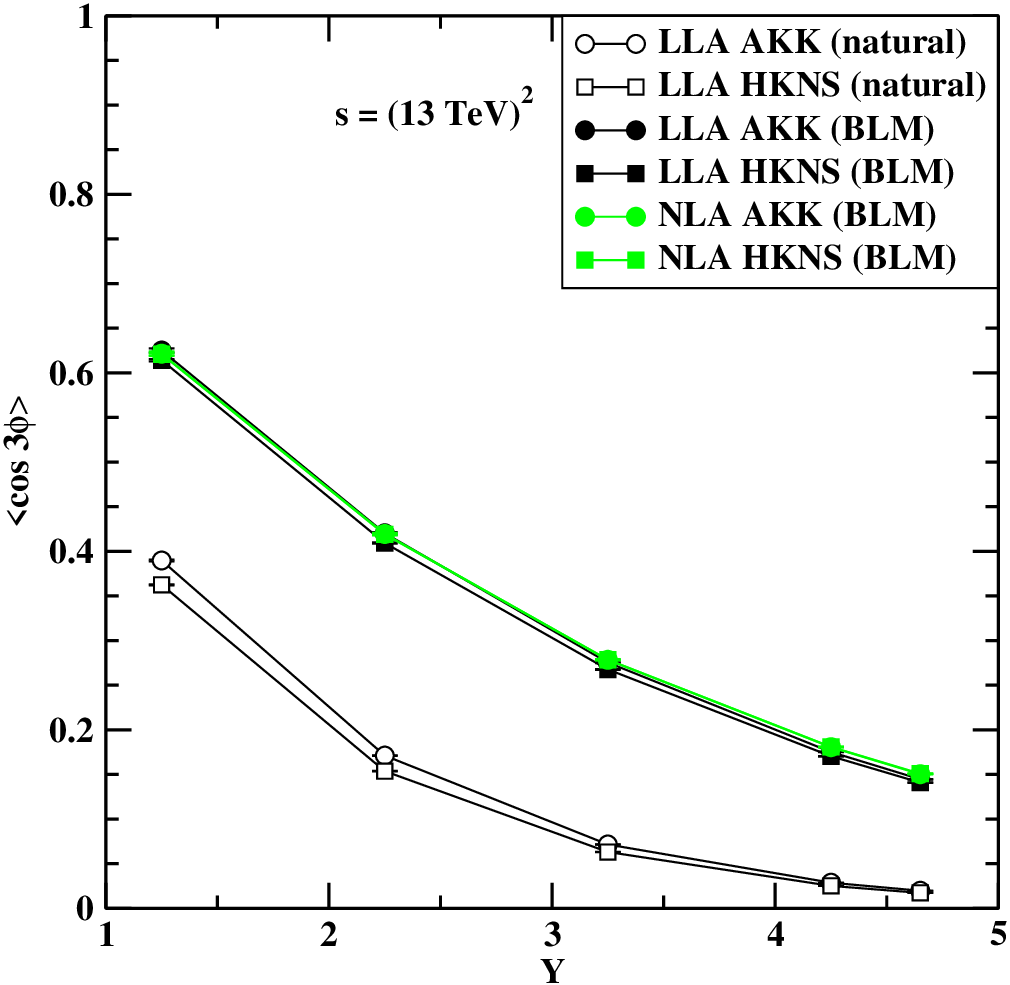}
\caption[]{$\langle \cos \phi \rangle$, $\langle \cos 2\phi \rangle$ 
and $\langle \cos 3\phi \rangle$ for di-hadron production 
at $\sqrt{s}=13$ TeV  for the two parametrizations of FFs considered in 
this work. See the text for the definition of ``natural'' and 
``BLM'' scales.}
\label{correlations}
\end{figure*}

\begin{figure*}[htb]
\centering
\includegraphics[scale=0.40,clip]{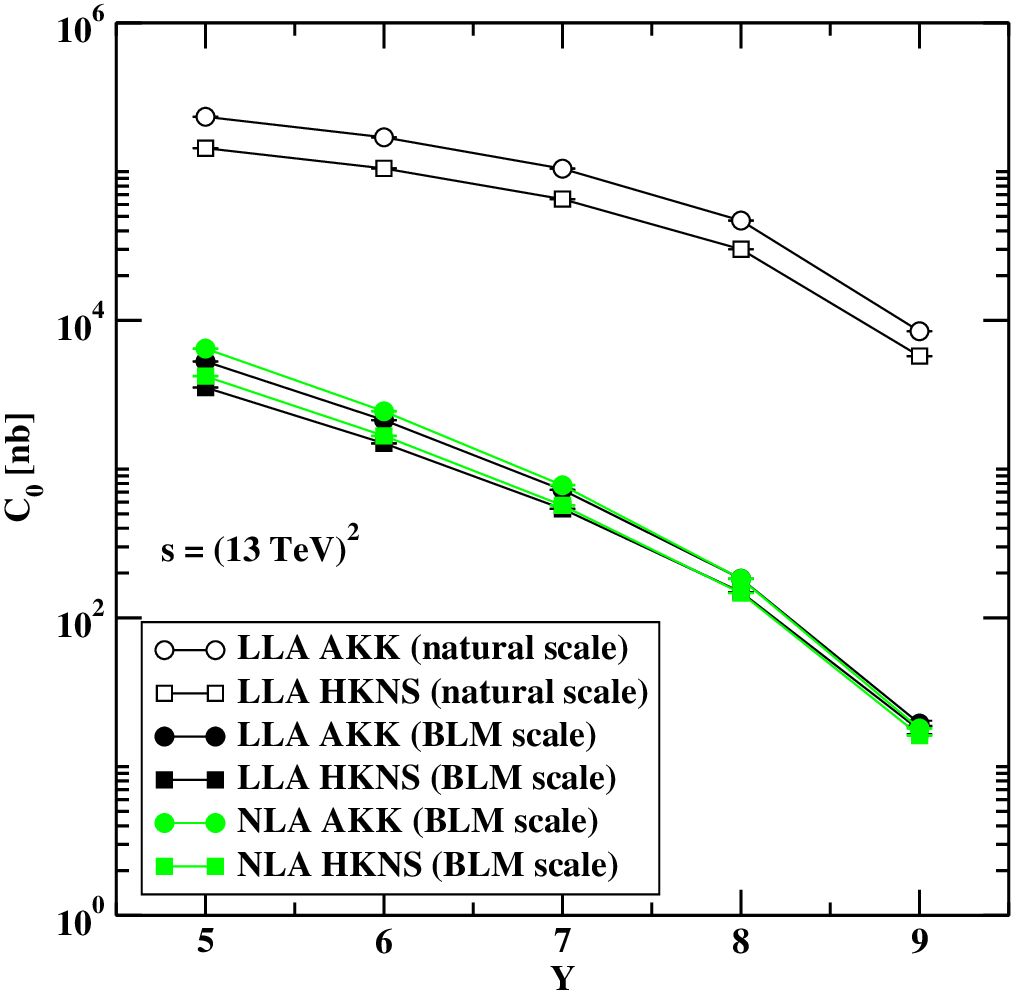} \ \ \
\includegraphics[scale=0.40,clip]{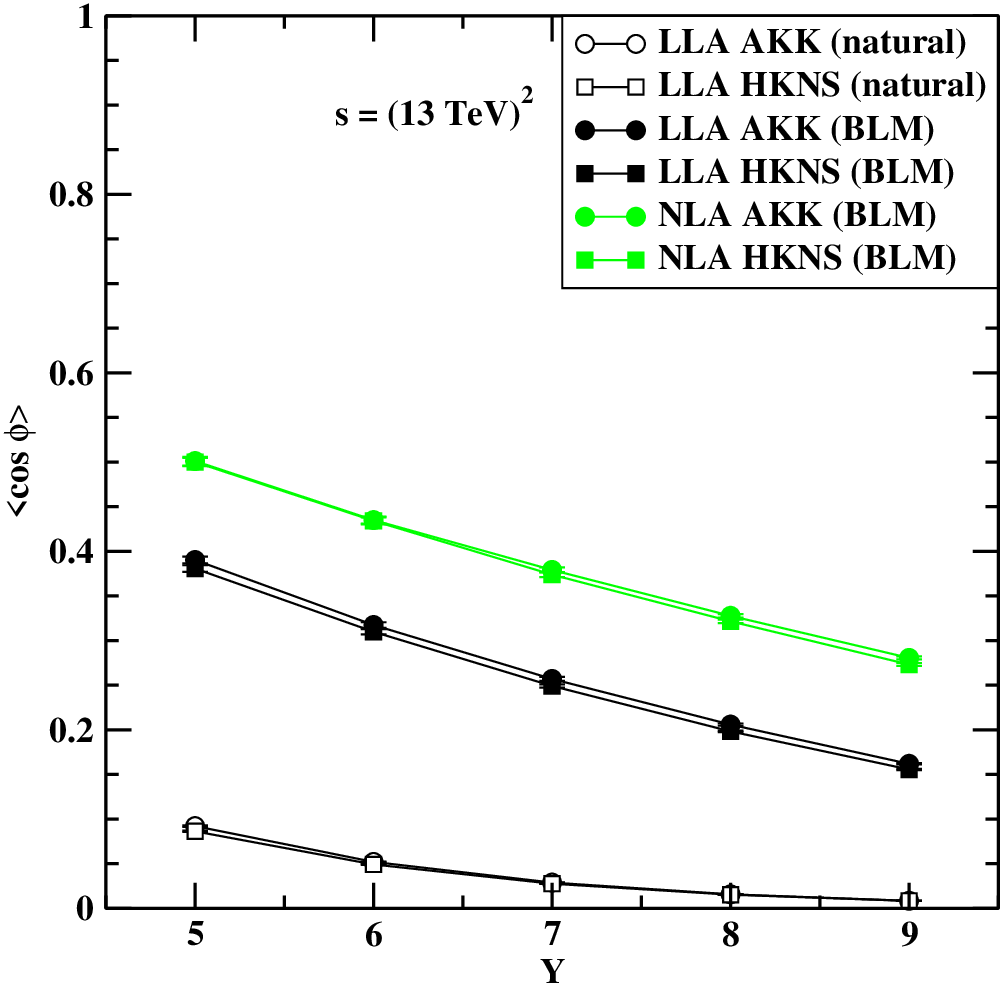}
\caption[]{Cross sections and $\langle \cos \phi \rangle$, for di-hadron production 
at $\sqrt{s}=13$ TeV and larger rapidity intervals $Y$.}
\label{largeY}
\end{figure*}

In order to find the values of the BLM scales, we introduce the ratios of the 
BLM to the ``natural'' scale suggested by the kinematic of the process, 
$\mu_N=\sqrt{|\vec k_{1}||\vec k_{2}|}$, so that $m_R=\mu_R^{\rm BLM}/\mu_N$, 
and look for the values of $m_R$ such that Eq.~(\ref{beta0}) is satisfied. 
Results are presented 
in Fig.~\ref{scales} as functions of $Y$ for the first few values of $n$ and 
for the two values of the LHC center-of-mass energy. Then we plug these 
scales into our expression for the integrated coefficients in the BLM scheme 
(for the derivation see~\cite{Caporale:2015uva}):
%\begin{widetext}
\beq{}
\label{eq}
C_n
=\int\PS  \,
\int\limits^{\infty}_{-\infty} d\nu \, \frac{e^Y}{s}\,
 e^{Y \bar \alpha^{\rm MOM}_s(\mu^{\rm BLM}_R)\left[\chi(n,\nu)
+\bar \alpha^{\rm MOM}_s(\mu^{\rm BLM}_R)\left(\bar \chi(n,\nu) +\frac{T^{\rm conf}}
{3}\chi(n,\nu)\right)\right]}
\eeq
\[
\times \left(\alpha^{\rm MOM}_s (\mu^{\rm BLM}_R)\right)^2 c_1(n,\nu)c_2(n,\nu)
\left[1+\bar \alpha^{\rm MOM}_s(\mu^{\rm BLM}_R)\left\{\frac{\bar c^{(1)}_1(n,\nu)}
{c_1(n,\nu)}+\frac{\bar c^{(1)}_2(n,\nu)}{c_2(n,\nu)}+\frac{2T^{\rm conf}}{3}
\right\} \right] \, .
\]
%\end{widetext}
The coefficient $C_0$ gives the total cross sections and the ratios
$C_n/C_0 = \langle\cos(n\phi)\rangle$ determine the values of the mean cosines,
or azimuthal correlations, of the produced hadrons. In Eq.~(\ref{eq}), 
$\bar \chi(n,\nu)$ is the eigenvalue of NLA BFKL kernel~\cite{Kotikov:2000pm}
 and its expression is given, {\it e.g.} in Eq.~(23) 
of~\cite{Caporale:2012ih}, whereas $\bar c^{(1)}_{1,2}$ are the NLA parts of 
the hadron vertices~\cite{hadrons}. The evaluation of~(\ref{eq}) requires a 
complicated 8-dimensional numerical integration (the expressions for 
$\bar c^{(1)}_{1,2}$ contain an additional longitudinal fraction integral in 
comparison to the formulas for the LLA vertices, given in~(\ref{c1}) 
and~(\ref{c2})).
Since the main aim of this work is to stress the potential relevance of 
the process we are proposing, rather than to give a full NLA prediction, 
we will present  our first results neglecting the NLA parts of hadron 
vertices, {\it i.e.} putting $\bar c^{(1)}_{1,2}=0$.  This  reduces the 
expression in the r.h.s. of Eq.~(\ref{eq}) to a 6-dimensional integral, 
manageable of a numerical calculation by a \textsc{Fortran} code.  
In Fig.~\ref{cross_sections} we present results for total cross sections 
at two values of the center-of-mass LHC energy: $\sqrt{s}=7$~TeV and 
$\sqrt{s}=13$~TeV. Fig.~\ref{correlations} shows our predictions for the 
azimuthal correlations at $\sqrt{s}=13$~TeV.   

For comparison, we considered also larger values of $Y$, similar 
to those used in the CMS Mueller-Navelet jets analysis. In 
Fig.~\ref{largeY} we present the cross section, ${\cal C}_0$, and 
$\langle \cos \phi \rangle$ in this larger $Y$-interval, calculated at 
center-of-mass energy of $\sqrt{s}= 13$ TeV and with the other settings as 
in Fig.~\ref{correlations}. These results may be of future reference for CMS, 
LHCb or other experiments.

\section{Discussion and outlook}

We checked that in our numerical analysis the essential values of $x$ are 
not too small, $x\sim [10^{-3}\div 10^{-2}]$, and even bigger in the case of the
larger $Y$ intervals presented in Fig.~\ref{largeY}. This justifies our use of 
PDFs with the standard DGLAP evolution. Note that our process is not a 
low-$x$ one, and similarly to the Mueller-Navelet jet production, we are 
dealing with a dilute partonic system. Therefore possible saturation effects 
are not important here, and the BFKL dynamics appears only through resummation 
effects in the hard scattering subprocesses, without influence on the 
PDFs evolution.

Our results are obtained using both the AKK and HKNS parameterizations 
for the hadron FFs. We see on Figs.~\ref{scales} and~\ref{cross_sections}
the sizeable difference between  predictions  in these two cases, which 
means that the FFs are not well constrained in the required kinematic region. 
In a similar range the difference between $\pi^{\pm}$ and $K^{\pm}$ AKK and 
HKNS FFs was discussed recently in~\cite{Yang:2015avi}.  Our calculation 
with the AKK FFs gives bigger cross sections, whereas the difference between 
AKK and HKNS in Fig.~\ref{correlations} is small, since the FFs uncertainties 
are largely  cancelled in the coefficient ratios describing the azimuthal angle
correlations.
We do not present separate plots for azimuthal correlations at $\sqrt{s}=7$~TeV 
because we found that the difference between our predictions for these 
observables at two LHC energies, $\sqrt{s}=7$~TeV and $\sqrt{s}=13$~TeV, 
is  not larger than $3\%$.   

The general features of our predictions for di-hadron production are rather 
similar to  those  obtained earlier for the Mueller-Navelet jet process. 
Although the BFKL resummation leads to the growth with energy of the partonic 
subprocess cross sections, the convolution of the latter with the proton PDFs 
makes the net effect of a decrease with $Y$ of our predictions in 
Fig.~\ref{cross_sections}.  This is due to the fact that, at larger values of 
$Y$, PDFs are probed effectively at larger values of $x$, where they
fall very fast. For the di-hadron azimuthal correlations we predict in 
Fig.~\ref{correlations}  a decreasing behavior  with $Y$. That originates 
from the increasing amount of hard undetected parton radiation in the final 
state allowed  by  the growth of the partonic subprocess energy. 
The values of the BLM scales we found are much larger than $\mu_N$, the  
scale suggested by the kinematic of the process. 
For the  BLM-to-natural  scale ratios we obtain rather large numbers,  
$m_R\sim 35$.  These  values are larger than those obtained previously for 
similar scale ratios in the case of the Mueller-Navelet jet production process. 
The difference may be attributed to the fact that, in the case of di-hadron 
production, we have an additional branching of the parton momenta (described 
by the detected hadron FFs), and typical transverse momenta of the partons 
participating in the hard scattering turn to be considerably larger than 
$|\vec k_{1,2}|$, the momenta of the hadrons detected in the final state.
We found that typical value of the fragmentation fraction is 
$z=\alpha_h/x\sim 0.4$, which explains the main source of the difference 
between the values of the BLM scales in the case of di-jet and di-hadron
production. Another source is related to the difference in the function 
$f(\nu)$, defined in~(\ref{nu}), which appears in the expression for the jet-
and hadron-vertex in these two reactions, and enters also the definition of 
the BLM scale: $f(\nu)$ is zero for the jets and non-zero in the di-hadron 
case.

Our predictions for di-hadron production calculated in LLA with the use of 
the natural scale $\mu_N$ and our NLA results obtained with the
BLM scale setting are different: with NLA BLM we got much lower values of the 
cross sections, see Fig.~\ref{cross_sections}, and considerably larger 
predictions for the $\langle \cos n\phi \rangle$, see Fig.~\ref{correlations}. 
For comparison, in Figs.~\ref{cross_sections} and~\ref{correlations} we show 
our NLA BLM predictions together with the results we 
obtained in LLA, but using the large values of the scales as  determined  
with the BLM setting.  Plots of Figs.~\ref{cross_sections} 
and~\ref{correlations} show that the LLA results 
with BLM scales lie closer to the NLA BLM ones than LLA results with 
 natural scales. The difference between NLA BLM and LLA with BLM scale 
predictions is due to the account of NLA corrections to the BFKL kernel 
in the former. In this paper we did not include the known results for the
NLA corrections to the hadron vertices, therefore our NLA analysis is 
 approximated. As the next stage, we plan to incorporate the terms  
$\bar c^{(1)}_{1,2}(n,\nu)$ in our numerical code. At present we can only rely 
on the experience gained with the analysis of the similar Mueller-Navelet jet 
production process, where it was shown that NLA effects coming from the 
corrections to the BFKL kernel and to the jet vertices are equally important. 
Therefore the difference between our incomplete NLA BLM and LLA with BLM scale 
results could be considered as a rough estimate of the uncertainty of the 
present analysis.  

The rapidity range we focused on here, $Y\leq 4.8$, may look to 
be not large enough for the dominance of BFKL dynamics. But we see, however, 
that in this range there are large NLA BFKL corrections, thus indicating that
the BFKL resummation is playing here a non-trivial role. To clarify the issue  
it would be very interesting to confront our predictions with the results of 
fixed-order NLO DGLAP calculations. But this would require new numerical 
analysis in our semihard kinematic range, because the existing NLO DGLAP 
results cover the hard kinematic range for the energies of fixed target 
experiments, see for instance~\cite{Owens:2001rr,Almeida:2009jt}.
 
In our calculation we  adopted, somewhat arbitrarily, the limit 
$|\vec k_{1,2}|\geq 5$~GeV for hadron transverse momenta. With this choice we 
obtained large values of the process cross sections, presented in 
Fig.~\ref{cross_sections}.  
This makes us confident that the inclusive production of two detected 
hadrons separated by large rapidity intervals  could be considered in
forthcoming analyses at the LHC. 

Considering a region of lower hadron transverse momenta, 
say $|\vec k_{1,2}| \geq 2$~GeV, would lead to even larger values of the cross sections. 
But it should be noted that in our calculation we use the BFKL method together 
with  leading-twist  collinear factorization, which means that we 
 are systematically neglecting power-suppressed corrections. 
Therefore, going to smaller transverse momenta we  would  enter a region 
where higher-twist effects must be important. The applicability border for our 
approach could be established either by comparing our predictions with future 
data or by confronting it with some other theoretical predictions which do  
include higher-twist effects. 
For the last point, one can consider an alternative, higher-twist production 
mechanism, related with multiparton interactions in QCD (for  a  review, 
see~\cite{Diehl:2011yj}). The double-parton scattering contribution to the  
Mueller-Navelet jet production was considered in the 
papers~\cite{Ducloue:2015jba} and~\cite{Maciula:2014pla}, using different 
approaches. It would be very interesting if similar estimates were done 
also for the case of di-hadron production.

In conclusion, we believe that, even within the approximation adopted 
in our calculation and the systematics uncertainties related with 
the scale setting procedure, we have provided enough evidence that
the study of the di-hadron production can be successfully included 
in the program of future analyses at the LHC and can improve our knowledge 
about the dynamics of strong interactions in the Regge limit. 

\section{Acknowledgments}

We thank G.~Safronov and I.~Khmelevskoi for stimulating and helpful discussions.
This work was supported in part by the RFBR-15-02-05868.

\end{document}